\documentclass[
aps,
prc,
reprint,
nofootinbib,
 amsmath,amssymb,
floatfix
]{revtex4-2}

\usepackage{graphicx}
\usepackage{hyperref}       
\usepackage[capitalize]{cleveref}
\crefname{section}{Sec.}{Sec.}
\Crefname{section}{Section}{Sections}

\usepackage{dcolumn}
\usepackage{bm}
\usepackage{xcolor}
\usepackage{enumitem}

\def\ccake{{\sc c\kern-.05em c\kern-.05em a\kern-.05em k\kern-.05em e\kern-.05em}}

\begin{document}

\title{Viscous Gubser flow with conserved charges to benchmark fluid simulations} 


\author{Kevin Ingles}
\affiliation{%
The Grainger College of Engineering, Illinois Center for Advanced Studies of the Universe, Department of Physics, University of Illinois at Urbana-Champaign, Urbana, IL 61801, USA
}%

\author{Jordi Salinas San Mart\'in}
\email{
jordis2@illinois.edu
}
\affiliation{%
The Grainger College of Engineering, Illinois Center for Advanced Studies of the Universe, Department of Physics, University of Illinois at Urbana-Champaign, Urbana, IL 61801, USA
}%

\author{Willian Serenone}
\affiliation{%
The Grainger College of Engineering, Illinois Center for Advanced Studies of the Universe, Department of Physics, University of Illinois at Urbana-Champaign, Urbana, IL 61801, USA
}%

\author{Jacquelyn Noronha-Hostler}
 \email{jnorhos@illinois.edu}
\affiliation{%
The Grainger College of Engineering, Illinois Center for Advanced Studies of the Universe, Department of Physics, University of Illinois at Urbana-Champaign, Urbana, IL 61801, USA
}%

%
%
\begin{abstract}
We present semi-analytical solutions for the evolution of both the temperature and chemical potentials for viscous Gubser flow with conserved charges.
Such a solution can be especially useful in testing numerical codes intended to simulate relativistic fluids with large chemical potentials.
The freeze-out hypersurface profiles for constant energy density are calculated, along with the corresponding normal vectors and presented as a new unit test for numerical codes.
We also compare the influence of the equation of state on the semi-analytical solutions.
We benchmark the newly developed Smoothed Particle Hydrodynamics (SPH) code \ccake{} that includes both shear viscosity and three conserved charges. 
The numerical solutions are in excellent agreement with the semi-analytical solution and also are able to accurately reproduce the hypersurface at freeze-out.
\end{abstract}

\maketitle
%
%
\section{Introduction}

Relativistic hydrodynamics has become a standard tool for studying the quark-gluon plasma, a state of matter believed to have existed immediately after the Big Bang~\cite{Heinz:2013th,Luzum:2013yya,DerradideSouza:2015kpt,Romatschke:2017ejr}.
The quark-gluon plasma is the hottest, densest, and most ideal fluid known, and it can be reproduced in heavy-ion collisions at high energies.
Critical to understanding the properties of the quark-gluon plasma have been comparisons between theoretical models based on relativistic viscous hydrodynamics and experimental data, performed on an event-by-event basis.  
To trust these comparisons, it is important that these hydrodynamic models undergo benchmark and convergence tests to verify their numerical accuracy. 
Such tests also determine simulation parameters---such as the grid size for grid-based codes or smoothing scale of particle-based codes---which affect the numerical accuracy at the expense of run time, i.e., efficiency. Exact and semi-analytical solutions for relativistic hydrodynamics provide such unit tests.

Two standard candles in this context are Bjorken flow~\cite{Bjorken:1982qr} and Gubser flow~\cite{Gubser:2010ze,Gubser:2010ui}, which offer analytical solutions to highly-symmetric systems. 
More recently, additional tests have been developed in Refs.~\cite{Shi:2022iyb,Bradley:2024jxq}.
These benchmarks have primarily been applied in the literature to hydrodynamic systems with vanishing chemical potential (i.e., no net-baryon number $B$, strangeness $S$, or electric charge $Q$).

Incorporating conserved currents (and consequently, chemical potentials) into a relativistic hydrodynamic framework is crucial for determining the critical point of the QCD phase diagram and studying neutron stars and their mergers~\cite{Baym:2017whm,Busza:2018rrf,Bzdak:2019pkr,Dexheimer:2020zzs,Monnai:2021kgu,An:2021wof,Lovato:2022vgq,Achenbach:2023pba,Sorensen:2023zkk,Du:2024wjm}.
There have been several investigations into how a finite chemical potential affects exactly solvable systems.
For instance, a study of phase diagram trajectories was preformed for Bjorken flow~\cite{Dore:2020jye,Dore:2022qyz,Chattopadhyay:2022sxk}, kinetic theory~\cite{Du:2020zqg}, and in a 3D-Ising model~\cite{Pradeep:2024cca}.
In Ref.~\cite{Denicol:2018wdp}, ideal Gubser flow with conserved charges was developed and subsequently employed to test a relativistic viscous hydrodynamic model and then tested against a code with $BSQ$ conserved charges in \cite{Plumberg:2024leb}.
A formulation of viscous Gubser flow with conserved charges was completed in Ref.~\cite{Du:2019obx}, however, the authors did not calculate the evolution of the chemical potentials.
Finally, viscous Gubser flow with conserved charges was applied in the Navier-Stokes limit to study flow coefficients in Ref.~\cite{Hatta:2015era}, using the
analysis in Ref.~\cite{Hatta:2014jva} at nonzero chemical potential.
However, the acausal structure inherent in Navier-Stokes theory renders it unsuitable for describing the early-time dynamics of heavy-ion collisions \cite{Hiscock:1985zz}.

In this paper, we present and study the evolution of the temperature and chemical potentials for viscous Gubser flow with conserved charges for the first time, thereby filling in a lingering gap in the literature.
We refer to this solution as viscous Gubser flow with conserved charges (VGCC)\footnote{The open source \textsc{python} implementation of this solution is available at \url{https://github.com/the-nuclear-confectionery/viscous-gubser-with-conserved-charges}.}.
The symmetries of Gubser flow fix the velocity fields and prohibit diffusion (see Appendix \ref{app:diffusion}), so the evolution equation for the conserved current $n_Y$, associated to a chemical potential $\mu_Y$, does not couple directly to that of the energy-momentum tensor.
Instead, the coupling of temperature and chemical potentials arises through the equation of state (EoS).
Here, we use two analytic equations of state to investigate the effects of including a chemical potential. Then, we show the freeze-out profile for VGCC, and compare the freeze-out hypersurface for systems with different chemical potentials to quantify the impact of the presence of conserved charges. 
We also present comparisons with the newly developed relativistic viscous hydrodynamic code with $BSQ$ conserved charges, \ccake{} \cite{Plumberg:2024leb}, which is based on the Smoothed Particle Hydrodynamics (SPH)~\cite{ccakesite} method.  
We benchmark \ccake{} against the semi-analytical solutions presented in this paper.
We find good numerical agreement between the semi-analytical and numerical solutions, with the largest discrepancy arising from the off-diagonal shear components.
Lastly, we compare the calculated freeze-out hypersurface from \ccake{} to the one obtained semi-analytically. 
These comparisons are presented as a new test for hydrodynamic codes.

The remainder of the paper is organized as follows.
In \cref{sec:gubser_flow}, we review the fundamentals of Gubser flow, introduce the equations of state used in this study, and present the equations of motion for the temperature, chemical potentials, and shear stress tensor components, given our EoS choices.
In \cref{sec:freeze-out}, we provide the details for defining a freeze-out hypersurface and its normal vectors. We present the semi-analytical solutions for different initial conditions in \cref{sec:semi_analytic_results}. Comparisons to \ccake{} are shown in \cref{sec:breakdown} for both the dynamical evolution and the freeze-out hypersurface. Finally, we conclude in \cref{sec:conclusion}.

%
%
\section{Gubser flow}\label{sec:gubser_flow}

Gubser flow is characterized by the symmetry group $SO(3)_q\otimes SU(1,1)\otimes \mathbb Z_2$, which describes transversely expanding and boost invariant solutions. 
Such a system is naturally described in Milne (hyperbolic) coordinates, with the metric
\begin{equation}
    ds^2 = - d\tau^2 + dr^2 + r^2 d\phi^2 + \tau^2 d\eta^2.
\label{eq:milne_metric}
\end{equation}
Here, $\tau = \sqrt{t^2 - z^2}$ is the proper time, $r = \sqrt{x^2 + y^2}$ the radius in the transverse plane, $\phi = \tan^{-1}(x/y)$ the azimuthal angle, and $\eta = \tanh^{-1}(z/t)$ the spacetime rapidity.
The symmetry group does not allow for magnetic field, diffusion, or bulk pressure, but does allow for shear stress in the transverse plane.
Gubser flow is defined by the analytic solution for the 4-velocity of the fluid~\cite{Gubser:2010ze,Gubser:2010ui}
\begin{subequations}
    \label{eq:milne velocity}
    \begin{align}
        u_\tau(\tau, r) &= - \cosh{\kappa(\tau, r)}, \\
        u_r(\tau, r) &= \sinh{\kappa(\tau, r)}, \\
        u_\phi(\tau, r) &= u_\eta(\tau, r) = 0,
    \end{align}
\end{subequations}
where $\kappa(\tau, r)$ is a function used to write the fluid velocity concisely and is given by
\begin{equation}
    \tanh{\kappa(\tau, r)} = \frac{2q^2 \tau r}{1 + q^2\tau^2 + q^2 r^2}.
\end{equation}
The parameter $q$ is an arbitrary, dimensionful constant with mass dimension 1 that we can set to $1\,\text{fm}^{-1}$  without loss of generality.
Note, however, that the evolution of the system does depend on the value of $q$.

\subsection{Hydrodynamics}

Conservation of energy and momentum is expressed as $D_\mu T^{\mu\nu} = 0$, where $D_\mu$ is the covariant 
derivative and $T^{\mu\nu}$ is the energy-momentum tensor.
From this energy-momentum  conservation law, we derive the equations of motion.
In a hydrodynamic system undergoing Gubser flow, the most general energy-momentum tensor in the Landau frame is given by \cite{Landau1987Fluid}
\begin{equation}
    T^{\mu\nu}
    =
    (\mathcal E + \mathcal P) u^\mu u^\nu + \mathcal P g^{\mu\nu} + \pi^{\mu\nu}\,,
\end{equation}
where $\mathcal E$ is the energy density, $\mathcal P$ is the thermal pressure and $\pi^{\mu\nu}$ is the symmetric, traceless shear stress tensor.
The corresponding equations of motion are obtained by projecting out the time-like and space-like components of the conservation law, and can be written as 
\begin{subequations}
    \label{eq:Tmunu-evol}
    \begin{align}
        D \mathcal E + (\mathcal E + \mathcal P)\theta + \pi^{\mu\nu} \sigma_{\mu\nu} &= 0\,,
        \label{eq:e-evol}
        \\
        (\mathcal E + \mathcal P)\Delta^\alpha\!_\nu D u^\nu + \nabla^\alpha \mathcal P + \Delta^\alpha\!_\nu D_\mu \pi^{\mu\nu} &= 0\,.
    \end{align}
\end{subequations}
Here, we have introduced the comoving derivative $D=u^\mu D_\mu$; the expansion rate $\theta = D_\mu u^\mu$; 
the velocity-shear tensor $\sigma^{\mu\nu} = \Delta^{\mu\nu\alpha\beta}D_{\alpha}u_\beta$ where $\Delta^{\mu\nu\alpha\beta} \equiv (\Delta^{\mu\alpha} \Delta^{\nu\beta} + \Delta^{\mu\beta} \Delta^{\nu\alpha})/2 - \Delta^{\mu\nu} \Delta^{\alpha\beta}/3$
is the symmetric, traceless projection operator with $\Delta_{\alpha\beta} = g_{\alpha\beta} + u_{\alpha}u_{\beta}$ as the spatial projection operator;
and the spatially projected covariant derivative $\nabla^\mu = \Delta^{\mu\nu}D_\nu$.
The first equation gives the time-evolution of the energy density.
The second equation gives the time-evolution of the fluid velocity $u^\mu$, which is fixed by the Gubser solution, and is, therefore, redundant.
The system is closed by supplying a relaxation-type equation for the viscous correction~\cite{Marrochio:2013wla,Baier:2007ix}
\begin{equation}
    \label{eq:pi-evol}
    \Delta^{\mu\nu}_{\alpha\beta} D\pi^{\alpha\beta} + \frac{\pi^{\mu\nu}}{\tau_R} + \frac{2\eta}{\tau_R}\sigma^{\mu\nu} + 
    \frac{4}{3} \pi^{\mu\nu}\theta = 0.
\end{equation}
Here, $\tau_R$ is the relaxation-time which controls how quickly the shear modes relax to their non-relativistic limit, and $\eta$ is the shear viscosity which describes the friction between sheets of fluid; the two are related by 
\begin{equation}
    \tau_R = \mathcal C \frac{\eta}{\mathcal E + \mathcal P}\,,
    \label{eq:tau-pi}
\end{equation}
where $\mathcal{C}$ is the relaxation-time constant which we choose to be 5.
We express $\eta$ as $\eta = \bar \eta s$, where $\bar \eta$ is a constant fixed to 0.2 and $s$ is the entropy density.
Note, for a system with no conserved currents, the dimensionless quantity $\tau_R T$, where $T$ is the temperature suitably defined, is the constant $\mathcal C \bar \eta$, since $\mathcal E + \mathcal P = s T$.
However, for a system with conserved charges, this is no longer true, as will be discussed below.

\subsection{Conserved currents}

A conserved current in a system with no diffusion is simply given by the constitutive relation
\begin{equation}
    N_Y^\mu = n_Y u^\mu\,,
\end{equation}
where $n_Y$ is the conserved charge density for the quantity $Y$. 
Its conservation is expressed as $D_\mu N^\mu_Y = 0$, leading to the equation of motion
\begin{equation}
    Dn_Y + n_Y \theta = 0.
    \label{eq:n-evol}
\end{equation}
Because the fluid velocity is fixed in Gubser flow, the evolution of the charge density is completely independent of dynamics of the energy-momentum tensor.
We associate to this conserved charge density a chemical potential $\mu_Y$.
Note that just as the dynamics of the conserved current decouple from the energy-momentum evolution, they also decouple from other conserved charges.

As a strongly interacting system, heavy-ion collisions have multiple conserved charges (baryon number $B$, strangeness $S$, and electric charge $Q$), which are, in general, coupled (see \cite{Monnai:2010qp,Greif:2017byw,Fotakis:2019nbq,Oliinychenko:2019zfk,Oliinychenko:2020cmr,Carzon:2023zfp,Monnai:2019hkn,Noronha-Hostler:2019ayj,Bellwied:2019pxh,Monnai:2021kgu,Karthein:2021nxe,Aryal:2020ocm,Denicol:2012cn,Almaalol:2022pjc,Fotakis:2022usk,Schafer:2021csj,Martinez:2019jbu,Carzon:2019qja} for various, recent studies exploring these effects). 
In Gubser flow, the equations of motion for multiple conserved charges are completely independent of each other and without diffusive terms:
\begin{subequations}\label{eq:density-evol}
    \begin{align}
        Dn_B + n_B \theta &= 0,\\
        Dn_S + n_S \theta &= 0,\\
        Dn_Q + n_Q \theta &= 0
    \end{align}
\end{subequations}
These equations can be solved independently by providing initial conditions.
However, the equation of state introduces a nontrivial mapping from the natural hydrodynamic variables $\left\{\mathcal E,n_B,n_S,n_Q\right\}$ into the natural equation of state variables $\left\{T,\mu_B,\mu_S,\mu_Q\right\}$ even for conformal theories (see, e.g., the discussion on fallback equations of state in \cite{Plumberg:2024leb}). 
Additionally, depending on the equation of state, there may be a strong dependence on $\mathcal E$ on the combination of $n_B,n_S$, and $n_Q$. 
Thus, the semi-analytical solution derived here can be solved for multiple conserved charges to test how well a hydrodynamic code can resolve a multidimensional equation of state as well.

\subsection{Equations of State}

To relate the time-evolution of the chemical potential $\mu_Y$ to the time-evolution of temperature and viscous corrections, we need an equation of state.
Typically, an equation of state is just the relationship between the pressure and energy or entropy densities, as well as charge densities. 
That is, $\mathcal P = \mathcal P(\mathcal E, n_B,n_S,n_Q)$ or $\mathcal P(s, n_B,n_S,n_Q)$. 
Alternatively, it can be cast in terms of the conjugate variables $\mathcal P = \mathcal P(T,\mu_B,\mu_S,\mu_Q)$.
In fact, the natural choice of variables for expressing an equation of state in the Grand Canonical Ensemble is that of $\left\{T,\mu_B,\mu_S,\mu_Q\right\}$, while hydrodynamic codes traditionally use $\left\{\mathcal E,n_B,n_S,n_Q\right\}$ or $\left\{s,n_B,n_S,n_Q\right\}$ (as in the case of the approach we will discuss in this work in \cref{sec:breakdown}).
Thus, we begin by establishing a mapping between
\begin{equation}\label{eqn:inversion}
    \mathcal P(T,\mu_B,\mu_S,\mu_Q) \leftrightarrow \mathcal P(s,n_B,n_S,n_Q),
\end{equation}
which we will refer to here has the \textit{inversion problem} for the equation of state. In the following, we define our equation of state as the relationship between the pressure and our natural variables: $\{T,\mu_B,\mu_S,\mu_Q\}$.
We will study two different possibilities for the equation of state $\mathcal P(T,\mu_B,\mu_S,\mu_Q)$.

For the first equation of state, we choose that of a massless quark-gluon plasma~\cite{10.1088/bk978-0-750-31060-4ch6}
\begin{equation}
    \textsc{EoS1}: \quad \mathcal P = g_\mathrm{QGP}\frac{\pi^2}{90}T^4 + g_q
    \left(\frac{\mu_Y^2 T^2}{216} + \frac{\mu_Y^4}{3888\pi^2}\right),
    \label{eq:eos1}
\end{equation}
where $Y\in\{B,S,Q\}$, $g_\mathrm{QGP} = 2(N_c^2 -1) + (7/2)N_c N_f$ is the degeneracy in the QGP, $N_c$ is the number of colors, $N_f$ the number of flavors, and $g_q = 4 N_c N_f$ is the degeneracy for the quarks only.
We choose as values, $N_c = 3.0$ and $N_f = 2.5$.

The second equation of state we consider is~\cite{Plumberg:2024leb}
\begin{equation}
    \textsc{EoS2}: \quad
    \mathcal P
    =
    g_\text{QGP}\frac{\pi^2}{90}T^4
    \left[
        1
        +
        \sum_Y
        \left(
            \frac{T_\ast}{\mu_{\ast}}
            \frac{\mu_Y}{T}
        \right)^2
    \right]^2,
    \label{eq:eos2}
\end{equation}
where $T_\ast$ and $\mu_{\ast}$ are constants and $Y\in\{B,S,Q\}$.
The ratio of $T_\ast/\mu_\ast = 1$ was set arbitrarily.
In principle, one could choose a different value that would change the overall value of the pressure but would not change the qualitative behavior.

Once the form of $\mathcal P(T,\mu_B,\mu_S,\mu_Q)$ is determined, we can calculate all remaining thermodynamic variables in the following manner. 
The charge density for charge $Y$ and the entropy are obtained by taking partial derivatives with respect to $\mu_Y$ and $T$ as
\begin{equation} \label{eq:thermo-relations}
    n_Y = \left(\frac{\partial \mathcal P}{\partial\mu_Y}\right)_{T,\mu_{X\neq Y}}, \qquad
    s = \left(\frac{\partial \mathcal P}{\partial T}\right)_{\mu_{Y}},
\end{equation}
where the subscripts on the right-hand side denote variables that are being held constant. It is implied the $Y$ stands for any conserved charge and that $X\not = Y$ implies all other conserved charges $X$ besides the one being considered ($Y$).
For example, if one calculates $n_B$ in a system that conserves $BSQ$, then one must take the derivative of the pressure with respect to $\mu_B$ at fixed $T$, $\mu_S$, and $\mu_Q$.

A very useful formula that expresses the relationship between the thermodynamic variables is the Gibbs-Duhem relation,
\begin{equation}
    \label{eq:gibbs_duhem}
    \mathcal E+\mathcal P = sT+\sum_Y n_Y\mu_Y,
\end{equation}
which can be used solve for either $s$ or $n_Y$ without taking derivatives. 
Given the conformal symmetry of Gubser flow, we can simplify the left-hand side of \cref{eq:gibbs_duhem}, using the conformal relation $\mathcal E = 3 \mathcal P$.
Note that while the pressure is reduced to a function of the energy density alone in the conformal case, it can still depend on $\{T,\mu_B,\mu_S,\mu_Q\}$.
The entropy density is then, $ s = \left(4\mathcal P - \sum_Y \mu_Y n_Y\right) / T$.

With \cref{eq:tau-pi,eq:gibbs_duhem}, it is now clear that the dimensionless quantity $\tau_R T$ given by
\begin{equation}
    \tau_R T = \frac{\mathcal C \bar\eta}{1 + \sum_Y n_Y \mu_Y/(sT)},
\end{equation}
need not be constant, unlike the case without conserved charges.

\subsection{Evolution equations for \texorpdfstring{$T$}{T}, \texorpdfstring{$\mu_Y$}{μ}, and \texorpdfstring{$\pi^{\mu\nu}$}{π}}
\label{subsec:evolution_equations}

The system of partial differential equations defined by Eqs.~\eqref{eq:Tmunu-evol}, \eqref{eq:pi-evol}, and \eqref{eq:n-evol}, can be transformed into a system of ordinary differential equations, by coordinate-transforming to de Sitter space, with the metric 
\begin{equation}
    d\hat s^2 = -d\rho^2 + \cosh^2{\rho}\left(d\theta^2 + \sin^2{\theta}d\phi^2\right) + d\eta^2.
\end{equation}
This is done by first Weyl rescaling the Milne metric in \cref{eq:milne_metric} by $\tau^2$, $d\hat s^2 = d s^2 / \tau^2$, and applying the coordinate transformation
\begin{subequations}\label{eq:milne-de Sitter relations}
    \begin{align}
        \sinh{\rho} &= -\frac{1 - q^2 \tau^2 + q^2 r^2}{2q\tau}, \\
        \tan{\theta} &= \frac{2qr}{1+q^2 \tau^2 - q^2 r^2}.
    \end{align}
\end{subequations}
From now on, we will use hatted variables to represent quantities in de Sitter coordinates and un-hatted variables to represented quantities in Milne coordinates.
The transformation defined by the Weyl rescaling and \cref{eq:milne-de Sitter relations} is useful to describe the expanding fluid in hyperbolic coordinates as a fluid with a static velocity profile in de Sitter space, namely, $\hat{u}_\mu=(-1,0,0,0)$.
The Milne quantities can be recovered by using the transformation laws\footnote{The parameter that determines the scale is the proper time $\tau$. The scaling dimension, or conformal weight, is determined by $m-n-k$, where $m$ the number of indices up, $n$ the number of indices down, and $k$ is the mass dimension~\cite{Loganayagam:2008is}. Special exceptions are the metric, 4-divergence, and 4-velocity.}
\begin{subequations}
    \label{eq:milne-soln-transform}
    \begin{align}
        \label{eq:velocity transform}
        u_\mu(\tau, r) &= \tau \frac{\partial \hat x^\nu}{\partial x^\mu}\hat u_\nu(\rho), \\
        \label{eq:temperature transform}
        T(\tau, r) &= \frac{\hat T(\rho)}{\tau}, \\
        \label{eq:chemical potential transform}
        \mu_Y(\tau, r) &= \frac{\hat \mu_Y(\rho)}{\tau}, \\
        \label{eq:energy transform}
        \mathcal{E}(\tau, r) &= \frac{\hat{\mathcal{E}}(\rho)}{\tau^4}, \\
        \label{eq:shear transform}
        \pi_{\mu\nu}(\tau, r) &= \frac{1}{\tau^2} \frac{\partial\hat x^\alpha}{\partial x^\mu}\frac{\partial \hat x^\beta}{\partial x^\nu}\hat \pi_{\alpha\beta}(\rho),
    \end{align}
\end{subequations}
where the power of $\tau$ ensures that the conformal dimensions of the two sides of the equation agree.
We give explicit formulas for the shear stress components in \cref{app:pi-transform}.

The system of nontrivial differential equations obtained from Eqs.~\eqref{eq:Tmunu-evol}, \eqref{eq:pi-evol}, and \eqref{eq:density-evol} that correspond to \textsc{EoS1} in \cref{eq:eos1} is 
\begin{subequations}
    \begin{align}
        \frac{1}{\hat T}\frac{d\hat T}{d\rho}
        &=
        -\frac{2}{3}\left[
            1
            -
            \left(
                1
                +
                \frac{1}{3\pi^2} 
                \frac{\hat \mu_Y^2}{\hat T^2}
            \right)
            f(\hat \mu_Y, \hat T)
            \hat{\bar \pi}
        \right]\tanh\rho,
        \\
        \frac{1}{\hat\mu_Y}\frac{d\hat \mu_Y}{d\rho}
        &=
        -\frac{2}{3}
        \left[
            1 
            +
            2f(\hat \mu_Y, \hat T)\hat{\bar \pi}
        \right]\tanh\rho,
        \\
        \frac{d\hat{\bar \pi}}{d\rho}
        &=
        \frac{4}{3\mathcal C}\tanh{\rho}
        -
        \frac{\hat{\bar\pi}}{\tau_R}
        -
         \frac{4}{3}\hat{\bar \pi}^2\tanh{\rho},
        \label{eq:eos1-shear}
    \end{align}
    \label{eq:eos1-evol}
\end{subequations}
where the function $f(\hat \mu_Y, \hat T)$ is defined by
\begin{multline}
    f(\hat \mu_Y, \hat T)
    =
    \\
    \frac{
        \left[
            216\pi^4 g_\mathrm{QGP}\hat T^4 
            +
            5g_q\left(
                18\pi^2 \hat \mu_Y^2 \hat T^2
                +
                \hat \mu_Y^4
            \right)
        \right] 
    }{
        144\pi^2 g_\mathrm{QGP} \hat T^2 (3\pi^2 \hat T^2 + \hat \mu_Y^2)
        -
        10g_q\hat \mu_Y^2(9\pi^2 \hat T^2 - \hat \mu_Y^2)
    }.
\end{multline}
Here, we have defined $\hat \pi \equiv \hat \pi^\eta_\eta$ and $\hat{\bar \pi} \equiv \hat \pi / (\hat{\mathcal E} + \hat{\mathcal P})$ as the reduced shear stress component, and used that $\hat \theta = -3\hat\sigma^\eta_\eta = 2\tanh\rho$; in addition, we have utilized the orthogonality and tracelessness properties of the shear stress tensor.

On the other hand, the corresponding system for \textsc{EoS2} in \cref{eq:eos2} is given by
\begin{subequations}
\label{eq:eos2-evol}
\begin{align}
    \frac{1}{\hat T} \frac{d \hat T}{d \rho}
    &=
    \left[
        \frac{1}{3}
        \left(
            \hat{\bar \pi} - 2
        \right)
        +
        \hat{\bar \pi}
        \sum_Y
        \left(\frac{T_\ast}{\mu_\ast}\frac{\hat\mu_Y}{\hat T}\right)^2
    \right]
    \tanh\rho\,,
    \\
    \frac{1}{\hat \mu_Y}
    \frac{d \hat\mu_Y}{d\rho}
    &=
    -\frac{2}{3}
    (1 + \hat{\bar \pi})\tanh \rho.
    \label{eq:eos1-evol-mu}
\end{align}
\end{subequations}
with the same evolution equation for the reduced shear stress component as in \cref{eq:eos1-shear}. 

In this paper, we will use \textsc{EoS1} in \cref{eq:eos1} to demonstrate the features of VGCC and \textsc{EoS2} in \cref{eq:eos2} to perform comparisons with the \ccake{} hydrodynamics code. 
While only one equation of state is needed to benchmark a simulation, we have chosen two different conformal equations of state to demonstrate the generality of the solution.

%
%
\section{Freeze-out Surfaces}\label{sec:freeze-out}

There are various different approaches to hadronization from relativistic viscous fluid dynamics.  
However, the most common approaches include either isothermal freeze-out or fixed energy density criteria. 
The freeze-out hypersurface for Gubser flow can be defined by requiring the energy density be equal to some threshold $\mathcal E_\mathrm{FO}$, i.e,
\begin{equation}
    \mathcal E(\tau_\mathrm{FO}, r) = \mathcal E_\mathrm{FO},
    \label{eq:freeze-out condition}
\end{equation}
which is chosen because it naturally captures a decreasing freeze-out temperature with increasing chemical potential~\cite{Alba:2014eba,Bellwied:2015rza,STAR:2017sal,Bellwied:2018tkc}.
This constraint allows us to define the freeze-out time as a function of the radial distance $\tau_\mathrm{FO} = \tau_\mathrm{FO}(r)$.
If we rewrite the freeze-out constraint as
\begin{equation}
    F(\tau, r) = \mathcal E_\mathrm{FO} - \mathcal E(\tau, r),
    \label{eq:freeze-out condition 2}
\end{equation}
then the normal vectors for the freeze-out surfaces are defined by 
\begin{equation}
    n_\mu
    =
    \partial_\mu F(\tau, r) 
    =
    \frac{\partial F}{\partial \tau} \hat{\bm \tau}
    +
    \frac{\partial F}{\partial r} \hat{\bm r},
    \label{eq:normals}
\end{equation}
where the bold-face, hatted variables denote unit vectors.
The normalized normal vectors are then (recall that we work in the mostly plus
convention)
\begin{equation}
    \hat n_\mu 
    =
    \frac{n_\mu}{\sqrt{-n_\mu n^\mu}}.
\end{equation}
Given the analytic formulas for the equation of state and the solution for the temperature and chemical potentials' evolution, 
we can numerically find the zeros of \cref{eq:freeze-out condition 2} and the normal vectors, \cref{eq:normals}.
In grid-based codes, obtaining the normal vectors for the hypersurface can be challenging and is typically accomplished by using the Cornelius algorithm \cite{Huovinen:2012is}. 
In SPH codes, the normal vectors can be determined analytically \cite{Plumberg:2024leb}. 
However, even in the SPH case where analytical solutions are possible for the normal vectors, numerical error can occur due to the hydrodynamic evolution itself and/or the time steps, $\delta \tau$ since a fluid cell rarely hits exactly $\mathcal E_\mathrm{FO}$ at a given point in time. 
Rather one sees the fluid cell pass by $\mathcal E_\mathrm{FO}$ and then chooses the closest time step where the energy density reproduces $\mathcal E_\mathrm{FO}$.
Thus, computation of normal vectors provides an additional test for hydrodynamic codes.

%
%
\section{Exact Solutions}\label{sec:semi_analytic_results}

\subsection{Dynamical evolution of Gubser flow}
\begin{figure*}[ht]
    \centering
    \includegraphics[width=0.70\textwidth]{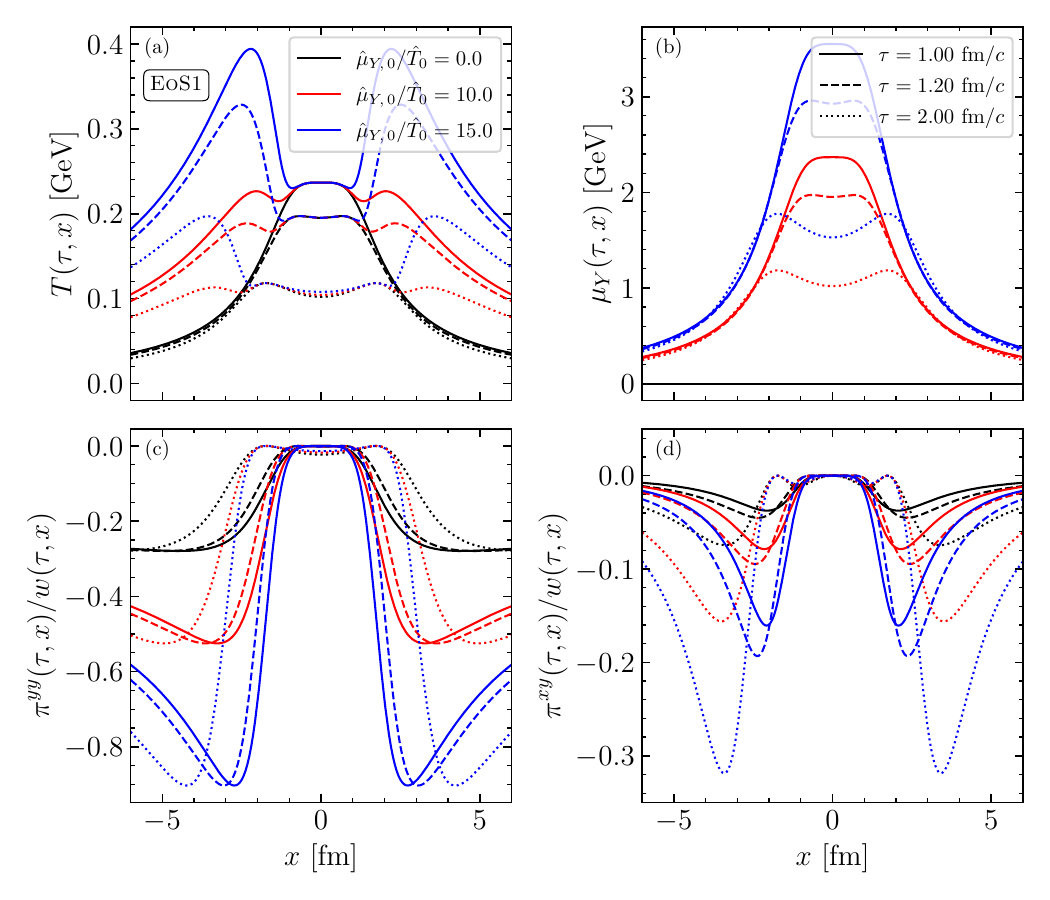}
    \caption{
    (Color online) We plot the semi-analytical solutions for Gubser flow with conserved charges in the plane $y=0$ (except $\pi^{xy}$, which is plotted in the plane $y=x$).
    Included are (a) the temperature, (b) chemical potential, (c) the dimensionless quantity $\pi^{yy}/w$, where $w=\mathcal E + \mathcal P$ is the enthalpy density, for a diagonal entry of the shear stress tensor, and (d) the dimensionless quantity $\pi^{xy}/w$ for an off-diagonal entry.
    The three different colors correspond to different initial conditions for the ratio of initial temperature
    and initial chemical potential: $\hat{\mu}_{Y,0}/\hat{T}_0=0$ (black), $10$ (red), $15$ (blue).
    The different types of lines correspond to selected times $\tau=1.0$ fm/$c$ (solid lines), $1.2$ fm/$c$ (dashed lines) and $2.0$ fm/$c$ (dotted lines) during the evolution.
    See text for discussion.
    }
    \label{fig:evolution comparison}
\end{figure*}

To investigate the semi-analytical solutions of the equations of motion in \cref{subsec:evolution_equations}, we use the initial conditions
\begin{subequations}
    \begin{align}
        \hat T_0 &\equiv \hat{T}(\rho=0) = 1.2, \\
        \hat{\bar \pi}_0 &\equiv \hat{\bar \pi}(\rho = 0)= 0,
    \end{align}
\end{subequations}
with different $\hat{\mu}_{Y,0}/\hat{T}_0$ ratios and the equation of state \textsc{EoS1} defined in \cref{eq:eos1}.
The system, Eqs.~\eqref{eq:eos1-evol}, is solved numerically and transformed to Milne coordinates using the relations in Eqs.~\eqref{eq:milne-soln-transform}.
We plot these solutions, for the initial conditions described above, in Fig.~\ref{fig:evolution comparison}.
We observe that the inclusion of conserved charges increases the magnitude of shear stresses in the system, and push it further away
from equilibrium.
For very large ratios of $\hat \mu_{Y,0}/\hat T_0$, the temperature profile in \cref{fig:evolution comparison}(a) develops shoulders away from the core due to viscous effects.
This shoulder structure persists throughout the entire evolution.
The presence of these shoulders follows from the relationship between the energy density $\mathcal E$, number density $n_Y$, and entropy density $s$, provided the ensemble relation in \cref{eq:gibbs_duhem} with a single conserved charge.

\begin{figure*}[htb]
    \centering
    \includegraphics[width=0.99\textwidth]{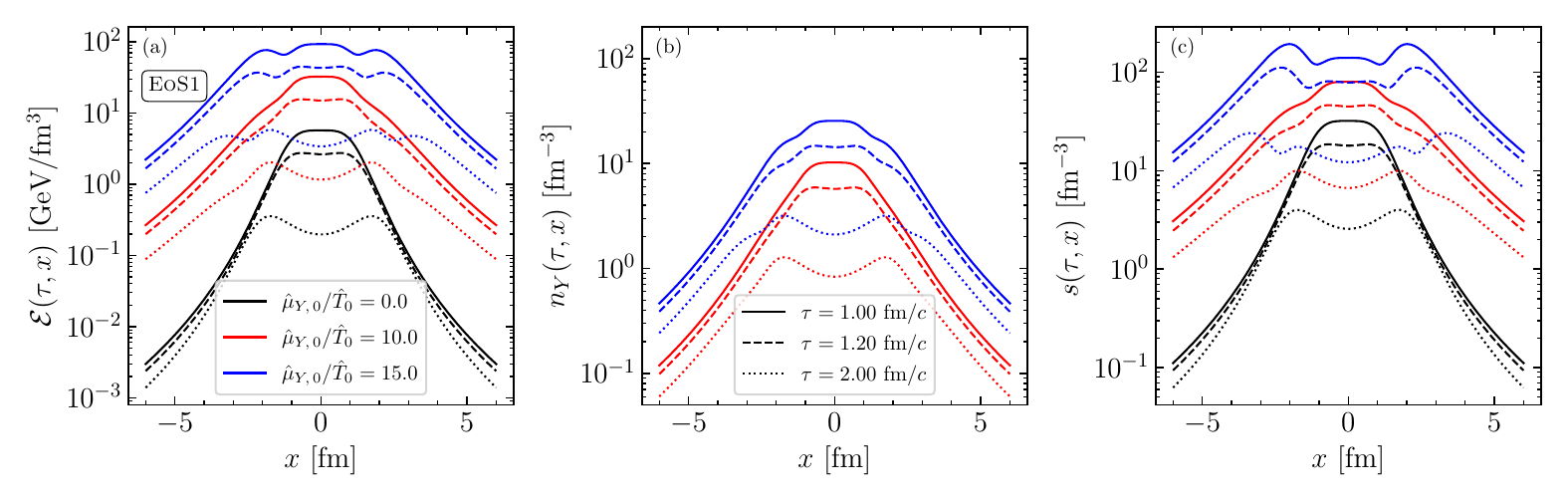}
    \caption{
        (Color online) We plot (a) the energy density, (b) number density, and (c) entropy density corresponding to the semi-analytical solutions for viscous Gubser flow with conserved charges shown in \cref{fig:evolution comparison} in the plane $y=0$.
        We include the results for differential values of the initial conditions and at different times (see the caption of \cref{fig:evolution comparison} for details).
        The number density plot does not feature black lines because at $\hat{\mu}_{Y,0}/\hat{T}_0=0$ the system has exactly vanishing number densities such that there is no evolution of the number density.
        Most notably, an increase in chemical potential can significantly decrease spatial variations of the energy
        density and entropy.
        The entropy density is most sensitive to the details of the temperature profile, as can be 
        seen by the presence of the shoulders for the blue lines ($\hat{\mu}_{Y,0}/\hat{T}_0 = 15$).
    }
    \label{fig:observables-1}
\end{figure*}

The energy density, number density, and entropy density, corresponding to the semi-analytical solution in Fig.~\ref{fig:evolution comparison}, have been plotted in Fig.~\ref{fig:observables-1}.
Due to the decoupled evolution of the number density in VGCC, the qualitative behavior of quickly decaying of away from the core is independent of equation of state.
Changing the initial condition $\hat\mu_{Y,0}/\hat T_0$ in the equation of state only results in shifting the profile of the number density in Fig.~\ref{fig:observables-1}(b) up or down in magnitude but does not change its  shape. 
Contrasting this to the equation of state and shear stress variables in Fig.~\ref{fig:evolution comparison}, we find that the shear entries and temperature profiles are significantly more sensitive to our choice in $\hat \mu_{Y,0}/ \hat T_0$.

We can also compare the spatial evolution of $\mu_Y$ in Fig.~ \ref{fig:evolution comparison}(b) to the spatial evolution of the number density $n_Y$ in Fig.~\ref{fig:observables-1}(b).
Away from the core the combined quantity $\mu_Y n_Y$ vanishes quickly such that $s T$ has to compensate to maintain the equality within the Gibbs-Duhem relation shown in Eq.~\eqref{eq:gibbs_duhem}, leading to the formation of shoulders
observed in the entropy density, and consequently, the temperature.
This increase in entropy, and consequently, increase of temperature is due to entropy production from the very large viscous corrections at finite chemical potential.
In the absence of a chemical potential these shoulders cannot form, regardless of how large $\hat{\bar{\pi}}_0$ is. 

%
%
\subsection{Freeze-out and entropy production}

\begin{figure*}[hb]
    \centering
    \includegraphics[width=0.99\textwidth]{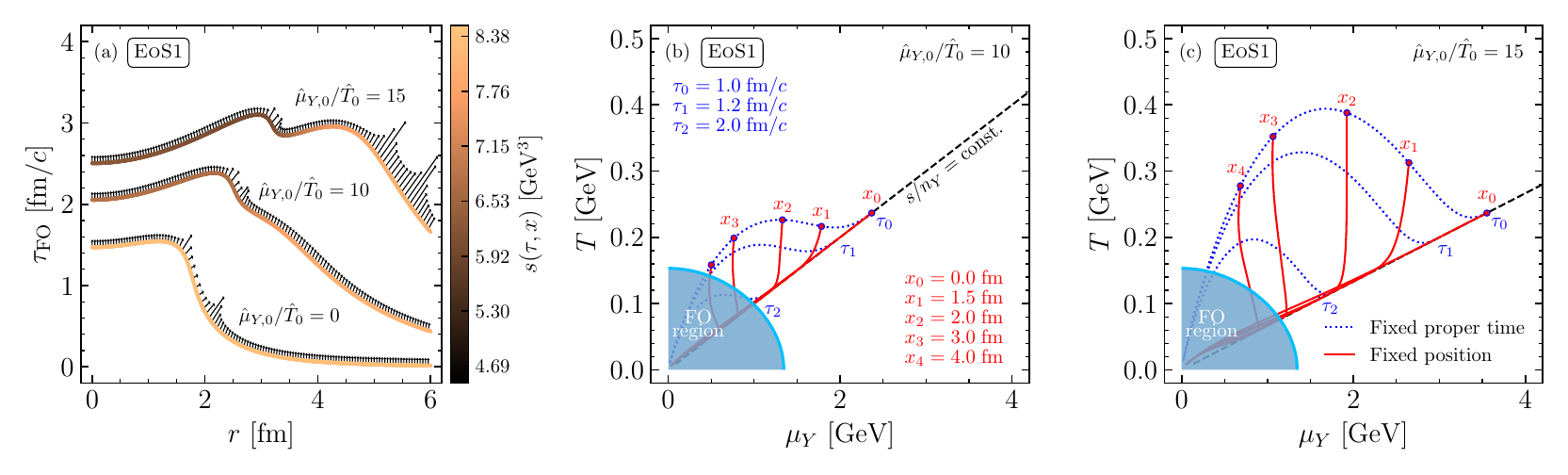}
    \caption{ 
        (Color online) We plot (a) the freeze-out hypersurface and normal vectors, and $T$ versus $\mu_Y$ trajectories for two initial conditions, (b) $\hat \mu_{Y,0}/ \hat T_0 = 10$ and (c) $\hat \mu_{Y,0}/ \hat T_0 = 15$ with $\hat T_0 = 1.2$.
        Included in the freeze-out hypersurface plot is a color bar which indicates the entropy density for a freeze-out cell at a given $(\tau_\mathrm{FO}, r)$.
        As expected, at zero chemical potential, the freeze-out hypersurface is isentropic.
        At non-vanishing chemical potential, the entropy in the core is lower (i.e., lower temperature) than the entropy at the tails.
        To allow the normal vectors and freeze-out hypersurface on the screen, they have been rescaled by a factor of $0.075$.
     }
    \label{fig:freeze-out surface}
\end{figure*}

The freeze-out hypersurfaces for different initial conditions $\hat\mu_{Y,0}/\hat T_0$ and freeze-out energy density $\mathcal E_\text{FO} = 1$ GeV/fm$^{3}$ are shown in \cref{fig:freeze-out surface}(a).
Conformal symmetry requires the energy density to be proportional to $T^4$.
At $\mu_Y=0$, this implies that the freeze-out hypersurface is isothermal and isentropic.
However, for $\mu_Y\neq 0$, freeze-out is neither isothermal nor isentropic.
Figure~\ref{fig:freeze-out surface}(a) shows that the entropy density (indicated by the color of the line and the color bar) close to the core is lower than the tails, i.e., freeze-out is no longer isentropic.
This follows, again, from the Gibbs-Duhem relation in \cref{eq:gibbs_duhem}: 
The left-hand side is fixed by the freeze-out condition, while on the right-hand side, since the $n_Y\mu_Y$ term vanishes at the edge of the system, the entropy (per conserved number density) has to increase to satisfy the equality.

The normal vectors to the freeze-out hypersurface are also presented in Fig.~\ref{fig:freeze-out surface}(a), where they have been rescaled by a factor of 0.075 to fit in the figure.
Causality requires all normal vectors to be future directed, which is satisfied in all the cases considered.
The divergences in normal vectors occur when the spatial and temporal variations of the energy density are close in magnitude.

In the middle and right panels of Fig.~\ref{fig:freeze-out surface}, we show the trajectories through the
$T$ versus $\mu_Y$ plane, for two initial conditions $\hat \mu_{Y,0} / \hat T_0 = 10$ (middle) and $\hat \mu_{Y,0} / \hat T_0 = 15$ (right).
The trajectories start at the top of the plot (on the fixed proper time surface with $\tau_0$ or fixed spatial coordinates $x$) and downwards in time.
Because the Gubser flow is composed of a field that spans multiple different values of $T-\mu_Y$ for a single set of initial conditions, we show a number of selected trajectories for different points in space.
Here, the blue dotted lines follow the profile of $(T,\mu_Y)$ at a fixed proper time $\tau$ and red solid lines are the trajectories along the phase diagram at a fixed point in space.
Notice how the shoulders seen for larger initial values $\hat \mu_{Y,0} / \hat T_0$ in the temperature profile of \cref{fig:evolution comparison}(a) enlarge the phase space probed (comparing the larger values and phase space in the right most figure vs the middle figure).
The blue shaded area marks the region where the energy density falls below the threshold $\mathcal{E}_\mathrm{FO}$, while the black dashed line represents the isentropic trajectory where $s/n_Y=\text{const.}$ and is fixed at freeze-out.
At early times the range of $T$ and $\mu_Y$ reached are to the left of the isentropic expansion, reaching lower chemical potentials $\mu_Y$ and either hotter or lower temperatures $T$, depending on the initial conditions. 
In addition, we also study the $(T,\mu_Y)$ evolution of a given point in the fluid, i.e., the trajectories at constant $x$ position. 
Interestingly enough, at a fixed spatial position the fluid remains at a nearly fixed $\mu_Y$ but only cools over time.

These trajectories are dramatically different from those observed in Refs.~\cite{Karthein:2021nxe,Noronha-Hostler:2019ayj,Monnai:2021kgu,Monnai:2019hkn} for realistic equations of state based on lattice QCD matched to a hadron resonance gas, which have a negative slope close to the origin and bend in the opposite direction.
The slope of these realistic EoS trajectories can be connected to a softening of the EoS close to the phase transition such that the speed of sound $c_s^2(T)$ reaches a minimum or, in other words, $\mathcal P(\mathcal E)$ has a softer slope. 
Nonetheless, in a conformal system $\mathcal E= 3\mathcal P$ and $c_s^2=1/3$ such that no softening occurs at hadronization and there is no bend in the isentropic trajectories.
Instead, our isentropic trajectories start at high $T$, high $\mu_Y$ and then steadily decrease by in $T$ and $\mu_Y$ over time.

In an out-of-equilibrium, non-conformal system it has been previously shown that the deviation from isentropes ($s/n_B$), in terms of the passage through the $T-\mu_B$ plane, is significantly more complex due to higher-order terms in the hydrodynamic equations of motion and the complex interplay between shear and bulk viscosities \cite{Dore:2020jye,Dore:2022qyz,Chattopadhyay:2022sxk}.
Here, even though our equation of state is significantly more simplistic, we still find more complex behavior of the out-of-equilibrium fluid over time. 
It would be interesting to study the temporal vs spatial evolution of a realistic EoS at finite $\mu_Y$ in future work.

%
%
\section{Comparisons of \texorpdfstring{\ccake{}}{CCAKE} to the Gubser test}\label{sec:breakdown}

\begin{figure*}[t]
    \centering
    \includegraphics[width=0.99\textwidth]{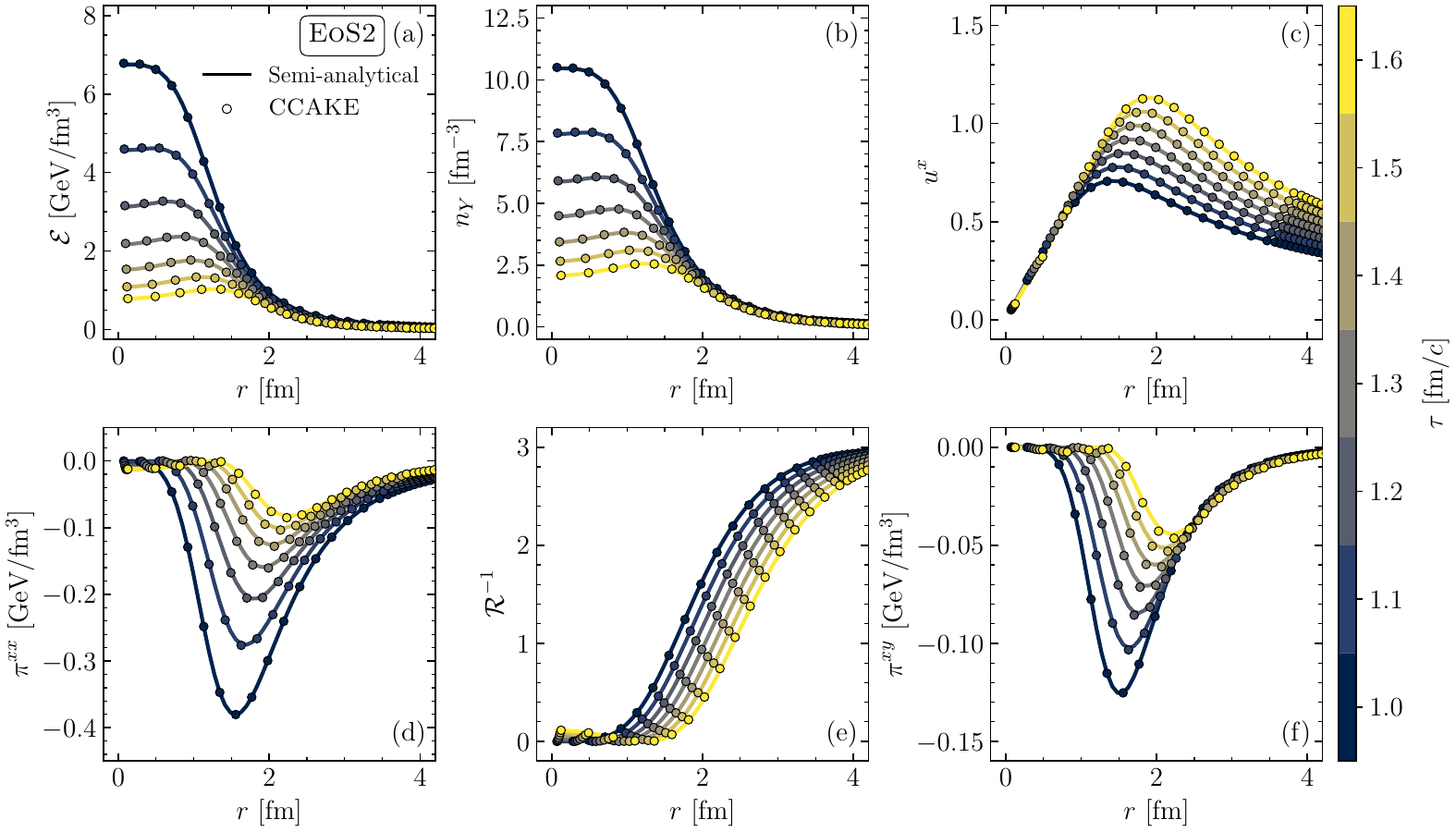}
    \caption{
        (Color online) We plot the numerical solution from the \ccake{} code (dots) and semi-analytical solution (solid line) 
        using \textsc{EoS2} in Eq.~\eqref{eq:eos2} as a function of distance from the origin and for various time steps (indicated by the color bar) within the first fm/$c$ of evolution.
        The comparison is made between (a) energy density, (b) number density, (c) the $x$-component of the fluid velocity, (d) the $xx$-component of the shear stress tensor, (e) 
        the inverse Reynolds number defined in Eq.~\eqref{eq:rey-pi}, and (f) the $xy$-component of the shear stress tensor.
        We see excellent agreement for all shown variables with a very mild deviation of around a percent around the core for the energy density and number density.
        For the $xx$-component of the shear stress tensor, we also observe an increasing deviation of the numerical solution from the semi-analytical one for increasing time, specifically between $r=1$ fm and $r=2$ fm, with the maximum relative deviations at $\tau = 1.6$ fm/$c$ getting large since the numerical solution is very close to zero. The largest deviations are seen on the $xy$-component of the shear stress tensor between $r=1$ fm and $r=2$ fm, and also present to a lesser extent on the diagonal component.
    }
    \label{fig:ccake-evol-comparison}
\end{figure*}

In the following section, we will use VGCC to benchmark the numerical solution of a hydrodynamic system with conserved charges implemented in the open source (2+1)-dimensional hydrodynamic code, Conserved Charge Hydrodynamik Evolution (\ccake~1.0.0)~\cite{Plumberg:2024leb}. \ccake{} relies on the SPH computational method, which discretizes the fluid into individual particles (known as SPH particles) and simplifies the calculation of terms in the equations of motion. 
The individual SPH particles move with the fluid such that one tracks their dynamical degrees of freedom (e.g., position, velocity, and entropy) over time. 
At each time step, a nearest neighbor search is used to evaluate the equation of motion. 
For the time-integration scheme, we use 2$^\text{nd}$-order Runge-Kutta (RK2), which has been tested against other methods and was found to optimize both accuracy and speed.
Previously, it was shown that \ccake~1.0.0 passed both the viscous Gubser test at vanishing densities and the ideal Gubser test at finite densities~\cite{Plumberg:2024leb}, but there was no test that combined both finite densities and viscosity, which is why we developed VGCC in this work. 

To compare with VGCC, we insert the semi-analytical solutions from \cref{eq:milne-soln-transform} with initial conditions
\begin{equation}
    \hat T_0 = 1.2, \quad \frac{\hat \mu_{Y,0}}{\hat T_0} = 0.3, \quad \hat{\bar \pi}_0 = 0,
\end{equation}
into \ccake{} and start our hydrodynamic evolution at $\tau_0 = 1.0\; \text{fm}/c$.
The grid size is $x, y \in [-5, 5]$ fm with a spacing of $0.05$ fm. 
We place an SPH particle on each grid point, which leads to a total of 40,401 SPH particles in each of these simulations.
The smoothing scale is set to $h=0.1$ fm to account for the large gradients in Gubser flow.
The equation of state used is \textsc{EoS2} and is given in \cref{eq:eos2}. 
Table~\ref{tab:sph_particles} summarizes the initial conditions, equation of state, and \ccake{} parameters used. The same VGCC parameters must be used in \ccake{} for an appropriate comparison. For instance, in VGCC we set the specific shear viscosity to $\bar{\eta}\equiv\eta/s=0.2$, which also must be used in \ccake{}.

\begin{table}[htbp!] 
\centering
\begin{tabular}{r@{\hspace{8pt}}l}
    \toprule
    \multicolumn{2}{c}{VGCC parameters}\\ 
    \colrule
        $\tau_0$ & 1.0 fm/$c$ \\
        $\hat{T}_0$ & 1.2 \\
        $\hat{\mu}_{Y,0}/\hat{T}_0$ & 0.3 \\
        $\hat{\bar \pi}_0$ & 0.0 \\
        $\mathcal{E}_\text{FO}$ & 1 GeV/fm$^3$ \\
        $\bar{\eta}=\eta/s$ & 0.20 \\
        $\tau_R$ & \cref{eq:tau-pi}  \\
    \toprule
    \multicolumn{2}{c}{EoS parameters}\\ 
    \colrule
        EoS & \textsc{EoS2} [\cref{eq:eos2}]  \\
        $\mu_\ast$ & 1 GeV \\
        $T_\ast$ & 1 GeV \\
    \toprule
    \multicolumn{2}{c}{\ccake{} parameters} \\ 
    \hline
        $h$ & 0.1 fm \\
        $\delta x$, $\delta y$ & 0.05 fm \\
        $\delta\tau$ & 0.001 fm/$c$ \\
    \botrule
\end{tabular}
    \caption{Parameters used to define the semi-analytical solutions for VGCC, equation of state choice and parameters (middle), and \ccake{} only parameters that determine the numerical accuracy (bottom) for comparisons. The same VGCC parameters and EoS must be used be used in both the semi-analytical solution and in \ccake{} for direct comparisons.
    }
    \label{tab:sph_particles}
\end{table}

Given the initial conditions discussed above, we compare the evolution calculated by \ccake{} with the solution predicted by our semi-analytical solution.
This comparison is presented in \cref{fig:ccake-evol-comparison} where we present the energy density $\mathcal E$,
a generic number density $n_Y$, $x$-component of the fluid velocity $u^x$, $xx$-component and $xy$-components of the shear stress tensor $\pi^{xx}$ and $\pi^{xy}$, and the inverse Reynolds number $\mathcal R^{-1}$ which we define as 
\cite[Eq.~8.10]{denicol2022microscopic}
\begin{equation}
    \mathcal{R}^{-1} = \frac{\sqrt{\pi^{\mu\nu}\pi_{\mu\nu}}}{\mathcal P}\,.
    \label{eq:rey-pi}
\end{equation}
We plot these quantities for multiple time steps as a function of the distance from the origin up to times of $\tau =1.6\; \text{fm}/c$, shown in Fig.\ \ref{fig:ccake-evol-comparison}.
Figure \ref{fig:ccake-evol-comparison} shows an excellent agreement overall between the semi-analytical and numerical solutions with minor deviations seen on $\pi^{xy}$ at late times at distances between 1 fm and 2 fm from the origin.
The simulations were run until they failed which is signaled by the presence of SPH particles with negative entropy. We reached failure at $\sim 1.85$ fm/$c$, as expected from previous numerical tests for \ccake{} \cite{Plumberg:2024leb} and its previous version v-USPhydro~\cite{Noronha-Hostler:2013gga,Noronha-Hostler:2014dqa}.
Note that unlike most grid-based codes, we do not include any regulator within SPH (for discussions on regulators see \cite{Shen:2014vra,Chiu:2021muk}).

The inverse Reynolds number, $\mathcal R^{-1}$, parametrizes how far from equilibrium a hydrodynamic system is.
Large $\mathcal R^{-1}$ indicates that the viscous contributions dominate and signals the breakdown of hydrodynamics (for examples of values regularly reached in heavy-ion collisions see \cite{Niemi:2014wta,Noronha-Hostler:2015coa,Summerfield:2021oex}).
Large $\mathcal{R}^{-1}$, in turn, leads to numerical instabilities which can cause the entropy density to become negative.
The size of the viscous corrections are also limited by causality \cite{Bemfica:2020xym,Plumberg:2021bme,Chiu:2021muk,ExTrEMe:2023nhy} and stability \cite{Almaalol:2022pjc} (that implies causality \cite{Gavassino:2021kjm}).
Causality violations are manifest at distances of $r\approx1.5–2.5\;\text{fm}$ away from the center and at late times, as demonstrated by \cref{fig:evolution comparison}(c).
In addition, causality violations are inherently linked to large $\mathcal R^{-1}$, hence we can conclude from \cref{fig:ccake-evol-comparison}(e) that the system at large radii (and early times) is likely well beyond causal constraints as well.
We do find that \ccake{} can correctly describe even very large values of $\mathcal{R}^{-1}$, but the eventual discrepancies and crashing of \ccake{} at late times is likely due to far-from-equilibrium behavior in other regions of the fluid.
Because \ccake{} is based on an SPH algorithm, the SPH particles move in time such that Fig.\ \ref{fig:ccake-evol-comparison} only shows their results in a snapshot in time, but they could have previously been in a part of the fluid that was far-from-equilibrium.
Thus, far-from-equilibrium behavior certainly leads to the breakdown of hydrodynamics but it is not clear yet where in the fluid this breakdown starts, only where it ends.

As noted, late times eventually lead to a crash in \ccake{} due to far-from-equilibrium and acausal behavior. 
A natural question that arises is if this concerning behavior appears before or after the conversion of fluid dynamics into a hadron resonance gas phase. 
The good news is that one can systematically check the freeze-out hypersurface against the semi-analytical solution derived from the Gubser test, which we do next. 
We find that all fluid cells freeze-out before \ccake{} crashes, implying that the problematic behavior only occurs at low temperatures, outside the regime where we expect hydrodynamics to be relevant. 

Within \ccake{}, the freeze-out hypersurface is calculated directly from the SPH particle information as one crosses the freeze-out criteria (either fixed temperature or energy density).  
However, as anticipated in Sec.~\ref{sec:freeze-out}, due to finite time steps $\delta \tau$ one almost never hits the freeze-out criteria \emph{exactly}, but is either slightly above or below; \ccake{} picks the result that minimizes the distance from the freeze-out criteria. 
With such an algorithm, the smaller the choice of $\delta \tau$, the more accurate the freeze-out hypersurface.

\begin{figure}[htbp]
    \centering
    \includegraphics[width=0.78\columnwidth]{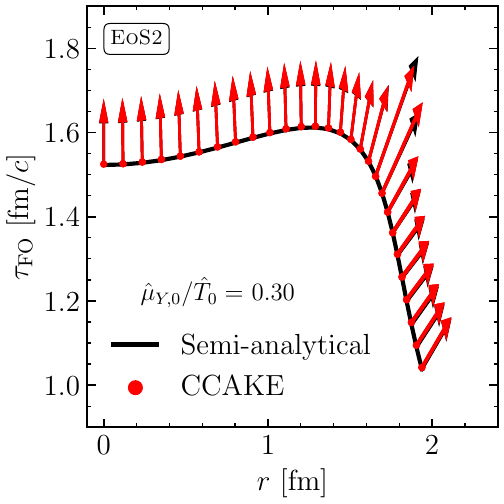}
    \caption{
        (Color online) A comparison between the freeze-out hypersurface and normal vectors obtained from \ccake{} (red) and the semi-analytical solution for VGCC (black). We see a slight disagreement on the magnitude of the (close to) divergent normal vectors.
        This disagreement is likely a result of the SPH particles freezing-out a time-step after the actual freeze-out time within the simulation.
    }
    \label{fig:ccake-FO-comparison}
\end{figure}

A comparison between the freeze-out hypersurface predicted by \ccake{} and the one calculated by VGCC is made in \cref{fig:ccake-FO-comparison}.
We see excellent agreement between the two.
Thus, this confirms that our approach to obtaining the hypersurface works well and is numerically accurate. 
Additionally, we find that for the Gubser test the breakdown of hydrodynamics occurs outside of the regime where we expect the system to behave as a fluid. 

\section{Conclusion}\label{sec:conclusion}

In this paper we provided the first study of the evolution of both the temperature and chemical potentials for viscous hydrodynamics systems with conserved currents undergoing Gubser flow.
We found that the addition of a chemical potential (as in contrast to $\mu_Y=0$) can push the fluid significantly further from equilibrium, dramatically changing the temperature and energy density profiles in the process. 
Our work allowed us to study out-of-equilibrium trajectories in a system with semi-analytical solutions to better understand its spacetime evolution.  
Compared to $\mu_Y=0$, we find that the $\mu_Y\neq 0$ evolution leads to particles at freeze-out that are significantly more boosted as indicated by the longer normal vectors on the freeze-out hypersurface.

The viscous Gubser flow with conserved charges (VGCC) solution provides a benchmark for hydrodynamics codes with conserved charges, such as \ccake{}.
We showed that \ccake{} passed our new Gubser test up until times of $\tau=\tau_\text{lim}\sim1.85$ fm/$c$.
At times after $\tau_\text{lim}$, \ccake{} crashes due to far-from-equilibrium effects that lead to negative entropy in certain SPH particles. 
However, we find that the fluid has already reached the freeze-out temperature by this time such that we can make direct comparisons between the VGCC and \ccake{} hypersurfaces. 
Lastly, we show that \ccake{} can very accurately reproduce the hypersurface generated using via Gubser flow. 

Further development of the ideas presented here would help bridge exact solutions and heavy-ion experimental results.
For example, the inclusion of a critical point on the EoS used would help elucidate exactly how a critical point affects the fluid in neighboring regions of the phase diagram.
Another interesting avenue is the implementation of a Gubser flow solution of relativistic viscous fluids in 3+1D with conserved charges, which would provide a helpful tool in the context of low and medium beam energies, where the chemical potentials probed are expected to be the largest.

\section*{Acknowledgements}
The authors would like to thank Kevin Pala, Surkhab Kaur Virk, and Jorge Noronha for useful discussions during the preparation of this manuscript.
This research was partly supported by the National Science Foundation under Grant No.  NSF PHYS-2316630, the US-DOE Nuclear Science Grant No. DE-SC0023861, and within the framework of the Saturated Glue (SURGE) Topical Theory Collaboration.
This work was supported in part by the National Science Foundation (NSF) within the framework of the MUSES collaboration, under grant number OAC-2103680.
Any opinions, findings, and conclusions or recommendations expressed in this material are
those of the author(s) and do not necessarily reflect the views of the National Science
Foundation.
We also acknowledge support from the Illinois Campus Cluster, a computing resource that is operated by the Illinois Campus Cluster Program (ICCP) in conjunction with the National Center for Supercomputing Applications (NCSA), which is supported by funds from the University of Illinois at Urbana-Champaign.

\bibliography{
bibliography/inspire,
bibliography/NOTinspire,
bibliography/supplementary
}

\appendix

\section{Vanishing diffusion in Gubser flow} \label{app:diffusion}
\subsection{Evolution of the diffusion current}
\noindent
For a generic conserved current, we can write
\begin{equation}
    N^\mu_Y=n_Yu^\mu + n_Y^\mu, \qquad Y\in \{B,S,Q\},
\end{equation}
where the first term is the equilibrium (ideal) contribution and the second, the diffusion current, encodes deviations from equilibrium. By construction, the diffusion current is orthogonal to the fluid velocity, $u_\mu n^\mu_Y=0$.

Following the Israel--Stewart formulation (or equivalently DNMR or related approaches that reduce to relativistic Navier--Stokes at lowest order), the diffusion current satisfies a relaxation-type equation \cite{Israel:1979wp,Monnai:2010qp,Denicol:2012cn,Fotakis:2019nbq,Almaalol:2022pjc,Fotakis:2022usk,Rocha:2023ilf},
\begin{equation}\label{eq:diffusion_current}
    \tau_{YZ}\Delta^{\mu\nu}\dot{n}_{\nu}^{Z}+n^\mu_Y = \kappa_{YZ} \Delta^{\mu\nu}D_\nu \left(\frac{\mu_{Z}}{T}\right),
\end{equation}
where $\tau$ and $\kappa$ denote the relaxation-time and diffusion coefficient matrices, respectively. The chemical potential associated with the conserved charge density $n_Y$ is $\mu_Y$ and $T$ is the temperature. Terms of order $\mathcal{O}(\text{Kn\,Re}^{-1})$, $\mathcal{O}(\text{Kn}^{2})$, $\mathcal{O}(\text{Re}^{-2})$, and higher are neglected. 

In de Sitter (hatted) coordinates introduced in \cref{subsec:evolution_equations}, \cref{eq:diffusion_current} becomes
\begin{equation}\label{eq:diffusion_rho_in_gubser_coordinates}
    \hat{n}^{\rho}_Y = 0,
\end{equation}
consistent with the condition of orthogonality to the fluid velocity, i.e., $\hat{u}_\rho \hat{n}^\rho_Y = \hat{n}^\rho_Y = 0$, and
\begin{subequations}
\label{eq:diffusion_current_in_gubser_coordinates}
\begin{align}
    \tau_{YZ}\left(\partial_\rho\hat{n}^\theta_{Z}+\tanh(\rho)\hat{n}^\theta_{Z}\right) + \hat{n}^\theta_{Y} &= 0, \\
    \tau_{YZ}\left(\partial_\rho\hat{n}^\phi_{Z}+\tanh(\rho)\hat{n}^\phi_{Z}\right) + \hat{n}^\phi_{Y} &= 0, \\
    \tau_{YZ}\partial_\rho\hat{n}^\eta_{Z} + \hat{n}^\eta_{Y} &= 0,
\end{align}    
\end{subequations}
where $\hat{\Delta}^{\mu}_{\:\,\nu}=\mathrm{diag}(0,1,1,1)$. The right-hand sides of \cref{eq:diffusion_current_in_gubser_coordinates} vanish because $SO(3)_q$ rotational invariance and $SO(1,1)$ boost invariance forbid $(\theta,\phi)$ and $\eta$ dependencies, respectively. Scalars, such as $\hat{\mu}_Y/\hat{T}$, may still depend on the time-like coordinate $\rho$, but this does not source diffusion since the projection operators remove the derivatives along $\hat{u}^\mu$.

Equation~(\ref{eq:diffusion_rho_in_gubser_coordinates}) shows that diffusion vanishes in the time-like direction, even when $D_\rho(\hat{\mu}_Y/\hat{T})\neq0$.
For all other directions, the homogeneous equations in \cref{eq:diffusion_current_in_gubser_coordinates} admit exponentially decaying solutions,
\begin{align}
    \hat{n}^\theta_Y(\rho) &= \frac{\cosh\rho}{\cosh\rho_0} \exp\left[-\left(\rho-\rho_0\right)\,\tau_{YZ}^{-1}\right]\hat{n}^\theta_{Z}(\rho_0),\\
    \hat{n}^\phi_Y(\rho) &= \frac{\cosh\rho}{\cosh\rho_0} \exp\left[-\left(\rho-\rho_0\right)\,\tau_{YZ}^{-1}\right]\hat{n}^\phi_{Z}(\rho_0),\\
    \hat{n}^\eta_Y(\rho) &= \exp\left[-\left(\rho-\rho_0\right)\,\tau_{YZ}^{-1}\right]\hat{n}^\eta_{Z}(\rho_0),
\end{align}
for seeded nonzero initial conditions and invertible $\tau$ matrices. Hence, if  $\hat{n}^\mu_Y(\rho_0)=0$ at some initial $\rho_0$, then it remains zero for all $\rho$.

\subsection{Constraints from exact Gubser symmetry}
\noindent
In an exactly Gubser-symmetric background, symmetry constraints forbid any nonzero diffusion current. This follows from the invariance requirements under the Gubser isometry group, as detailed below.
\subsubsection{\texorpdfstring{$SO(3)_q$}{SO(3)_q} rotational invariance}
\noindent
A vector field $V$ invariant under $SO(3)_q$, generated by the Killing vectors
\begin{subequations}\label{eq:Killing_vectors}
    \begin{align}
        L_x &= \sin\phi\,\partial_\theta + \cot\theta\cos\phi\,\partial_\phi,\\
        L_y &= -\cos\phi\,\partial_\theta + \cot\theta\sin\phi\, \partial_\phi,\\
        L_z &= \partial_\phi,
    \end{align}
\end{subequations}
corresponding to rotations about the $x$, $y$, and $z$ axes, must satisfy (see Ref.~\cite{Gubser:2010ze} for a more detailed explanation)
\begin{equation}\label{eq:Lie_derivative}
    \mathcal{L}_{L_i}V^a=L_i^b\partial_b V^a - V^b\partial_bL_i^a =0, \qquad a,b\in\{\theta,\phi\},
\end{equation}
where $\mathcal{L}_{L_i}$ is the Lie derivative along the Killing vector $L_i$ and $V$ is a tangent vector on the two-sphere $S^2$, i.e.,
\begin{equation}
    V = V^\theta(\rho)\,\partial_\theta + V^\phi(\rho)\,\partial_\phi.
\end{equation}
Since $V^a$ has no explicit $\theta,\phi$ dependence, the first term in \cref{eq:Lie_derivative} vanishes, and only the second term survives. Plugging in \cref{eq:Killing_vectors} into \cref{eq:Lie_derivative}, we reach
\begin{subequations}
    \begin{align}
        \mathcal{L}_{L_x}V^\theta &= - V^\phi\cos\phi,\\
        \mathcal{L}_{L_x}V^\phi &= V^\theta\csc^2\theta\cos\phi+V^\phi\cot\theta\sin\phi.
    \end{align}
\end{subequations}
Requiring these to vanish for all $(\theta,\phi)$ forces
\begin{equation}\label{eq:SO3_components}
    V^\theta=V^\phi=0.
\end{equation}
The same conclusion follows from $L_y$ and $L_z$. Hence $SO(3)_q$ symmetry forbids diffusion currents in the $\theta$ and $\phi$ directions.

\subsubsection{\texorpdfstring{$SO(1,1)\times \mathbb Z_2$}{SO(1,1) x Z_2}  invariance in rapidity}
\noindent
For a vector field along $\eta$, $V=V^\eta(\rho)\,\partial_\eta$.  
Boost invariance ($SO(1,1)$) means invariance under translations $\eta\mapsto\eta+\delta\eta$, i.e. $\partial_\eta V^\eta=0$. This allows a constant $V^\eta(\rho)$. However, $\mathbb Z_2$ reflection invariance requires $\eta\mapsto -\eta$. Under this transformation the basis vector flips sign,
\[
\partial_\eta \;\mapsto\; -\,\partial_\eta,
\]
so the field transforms as $V^\eta\partial_\eta \mapsto -V^\eta\partial_\eta$. Invariance under reflection requires $V^\eta=-V^\eta$, hence
\begin{equation}\label{eq:SO11_component}
V^\eta=0.
\end{equation}

Equations~(\ref{eq:diffusion_rho_in_gubser_coordinates}), (\ref{eq:SO3_components}), and (\ref{eq:SO11_component}) together imply that diffusion currents vanish identically in systems with exact Gubser symmetry.

\section{de Sitter-to-Milne transformations} \label{app:pi-transform}
\noindent
In this appendix, we present explicit expressions for the coordinate transformation of the shear stress tensor in Milne coordinates.
In the following, all quantities with hats are in de Sitter coordinates introduced in \cref{subsec:evolution_equations}, while un-hatted quantities are in Milne coordinates.
Recall that $\pi_{\mu\nu}$ is diagonal in de Sitter coordinates,
\begin{equation}
    \hat \pi_{\mu\nu}
    =
    -\frac{\hat \pi}{2}
    \mathrm{diag}(0,\cosh^2 \rho,\cosh^2\rho \sin^2\theta, -2).
\end{equation}
To transform these to Milne coordinates, we will need the
identities
\begin{subequations}
    \label{eq:coord-trans-dS-Milne}
    \begin{align}
        \left(
            \frac{\partial \theta}{\partial \tau}
        \right)^2
        \cosh^2{\rho} 
        &=
        \frac{u_r^2}{\tau^2},
        \\
        \frac{\partial \theta}{\partial \tau} 
        \frac{\partial \theta}{\partial r}\cosh^2{\rho} 
        &=
        \frac{1}{\tau^2}u_r u_\tau,
        \\
        \left[
            \left(
                \frac{\partial \theta}{\partial r}
            \right)^2
            -
            \frac{\sin^2\theta}{r^2}
        \right]
        \cosh^2\rho
        &=
        \frac{u_r^2}{\tau^2},
        \\
        \left[
            x^2\left(
                \frac{\partial \theta}{\partial r}
            \right)^2
            -
            \frac{y^2}{r^2}\sin^2\theta
        \right]
        \cosh^2\rho 
        &=
        \frac{r^2}{\tau^2}
        \left(
            1 
            +
            \frac{x^2}{r^2}
            u_r^2
        \right),
    \end{align}
\end{subequations}
where $\theta$ and $\rho$ are defined in Eqs.~\eqref{eq:milne-de Sitter relations} and $u_r$ and $u_\tau$ are given in Eqs.~\eqref{eq:milne velocity}.
Lastly, we note that $u_x = (x/r)u_r$. 
Therefore,
\begin{equation}
    \pi_{\mu\nu}
    =
    -\frac{\hat \pi}{2\tau^4}
    \begin{pmatrix}
        u_r^2 & u_x u_\tau & u_y u_\tau &   \\
              & 1 + u_x^2  & u_x u_y    &   \\
              &            & 1 + u_y^2  &   \\
              &            &            & -2\tau^2
    \end{pmatrix}.
\end{equation}


\end{document}